# Fluctuating exciton localisation in giant π-conjugated spoked-wheel macrocycles

Vikas Aggarwal[1], Alexander Thiessen[2], Alissa Idelson[1], Daniel Kalle[1], Dominik Würsch[3], Thomas Stangl[3], Florian Steiner[3], Stefan-S. Jester[1], Jan Vogelsang[3], Sigurd Höger[1,*)], John M. Lupton[2,3,*)]

[1]Kekulé-Institut für Organische Chemie und Biochemie der Universität Bonn, Gerhard-Domagk-Str. 1, 53121 Bonn, Germany

[2]Department of Physics and Astronomy, University of Utah, Salt Lake City, UT 84112, USA

[3]Institut für Experimentelle und Angewandte Physik, Universität Regensburg, D-93040 Regensburg, Germany

**Table of Contents Summary:**

We demonstrate the fundamentally non-deterministic nature of localisation of excitation energy in π-conjugated macromolecules for organic electronics. Shape-persistent ring structures mimic conjugated polymers in a structurally highly-ordered template. Single-molecule spectroscopy reveals intrinsic fluctuations in transition dipole orientation due to spontaneous symmetry breaking, leading to unpolarised fluorescence from individual molecules.

[*)] Corresponding authors. hoeger@uni-bonn.de, john.lupton@ur.de

**Conjugated polymers offer potential for many diverse applications but we still lack a fundamental microscopic understanding of their electronic structure. Elementary photoexcitations - excitons - span only a few nanometres of a molecule, which itself can extend over microns, and how their behaviour is affected by molecular dimensions is not fully understood. For example, where is the exciton formed within a conjugated segment, is it always situated on the same repeat units? Here, we introduce structurally-rigid molecular spoked wheels, 6 nanometres in diameter, as a model of extended pi-conjugation. Single-molecule fluorescence reveals random exciton localisation, leading to temporally-varying emission polarisation. Initially, this random localisation arises after every photon absorption event because of temperature independent spontaneous symmetry breaking. These fast fluctuations are slowed to millisecond timescales following prolonged illumination. Intramolecular heterogeneity is revealed in cryogenic spectroscopy by jumps in transition energy, however, emission polarisation can also switch without a spectral jump occurring, implying long-range homogeneity in local dielectric environment.**

Cyclic structures of varying levels of symmetry are ubiquitous in nature: from benzene and pyrrole over members of the porphin family such as haem or chlorophylls to photosynthetic antenna complexes[1], structural rigidity is crucial to molecules on different length scales. Yet synthetic compounds that derive macroscopic functions by mimicking elementary aspects of electron or energy transfer in organic semiconductors tend to be linear in structure[2]. While such materials, most notably π-conjugated polymers, possess a range of desirable functional characteristics, formulating a comprehensive microscopic picture of how individual covalently-bound monomer units arrange in space to form discrete π-conjugated segments remains challenging[3-6]. Conjugation and shape of the molecule are fundamentally interlinked[5]. On the one hand, spectroscopic techniques can, in principle, unveil information on electronic structure. On the other hand, physical shape, which can exhibit a level of diversity reminiscent of conformational degrees of freedom in proteins, is much harder to assess.

A conjugated polymer consists of a chain of repeat units. Π-electrons delocalise between monomers, but may become localised on longer scales due to the formation of chemical or structural defects[7]. An individual segment of the polymer which supports the π-orbital is referred to as a chromophore. The elementary excitation on a chromophore is an exciton – a tightly-bound electron-hole pair. Polymer optical properties such as spectral shape (width of energy bands and strength of vibronic coupling) are accounted for by excitonic coupling models where *intrachromophoric* interactions between monomers are described in the framework of *J*-aggregates, and interactions *between* chromophores are ascribed to either *J*- (inline) or *H*-aggregation (cofacial)[8]. The exciton itself is of order 2nm in size, which can be much smaller than the actual conjugated segment[7]. Depending on the magnitude of structural relaxation in the excited state, the exciton may be free to move within the chromophore, thereby increasing transition intensity since the number of electrons involved in the transition increases[8]; but it may also become localised[7]. Proximal chromophores can couple to each other, leading to further spreading of excitation energy in the macromolecule[9-10]. Fluorescence spectroscopy is often used to infer information on electronic structure, coupling mechanisms and conformation, but without definitive knowledge of molecular conformation to begin with, the parameters remain intractable. In particular, it is not self-evident that an exciton should always form on precisely the same monomer units of a chromophore. Structural relaxation in the excited state breaks molecular symmetry and leads to exciton self-trapping[11-12], the spatial localisation of excitation energy due to the nuclear rearrangement of the molecule following a redistribution in charge density. Does this process always follow the same pathway?

To examine these questions, we designed a giant molecular spoked-wheel structure with conjugated shape-persistent macrocyclic rim as a model of chromophore formation and interchromophoric coupling[13]. Using single-molecule techniques, we uncover two distinct localisation mechanisms: spontaneous symmetry breaking, with the exciton localising randomly to different parts of the ring after every photoexcitation event; and slower *photoinduced* symmetry breaking, leading to fluctuating exciton localisation on the millisecond timescale.

**Results and Discussions**

*Ring design*

The design of conjugated macrocyclic structures requires careful consideration of rigidity to prevent scissions in the π-system due to deformation of the overall ring as a result of the limited persistence length of rigid-rod building blocks[14]. We employ a phenylene-ethynylene-butadiynylene-based scaffold[15] to template carbazole units, offering an optimum in rigidity combined with desirable optical properties in the visible. Previously studied carbazole-based compounds had electronic transitions in the UV[16-18], making them poorly suited to single-molecule investigations, and porphyrin-based macrocycles of comparable rigidity to our rings showed very low fluorescence yields[19-25] with optical transitions in the near-IR. Other intriguing cyclic structures were not fully conjugated[26-27]. Figure 1 illustrates the design approach. Six *N*-phenyl-carbazole units are linked to each other by phenylene-ethynylene butadiynylene moieties and form the nominally-conjugated rim (green) of the spoked wheel **1**, where the spokes are not part of the rim conjugation. The ring is synthesised by Buchwald-Hartwig coupling of the rim segments **2** (green) and spoke modules **3** (red), selective removal of the cyanopropyldimethylsilyl protecting group from **4**[28], and coupling of the resulting acetylene to the hub **6** in a six-fold Sonogashira reaction. Subsequent deprotection of **7** leads to the spoked-wheel precursor **8**, which is cyclised in a palladium catalysed reaction under pseudo-high-dilution conditions and purified by recycling GPC to give **1** in 54% yield. The purity is confirmed by mass spectrometry, NMR and GPC (see *Supplementary Information*). We compared ring **1** to model oligomers **9**-**11**. The structural rigidity is visualised in the scanning-tunnelling micrographs (STM) in Fig. 2 (details see *Supplementary Information*). Although interactions with the hexagonal graphite substrate lattice and dense packing of the molecules induce slight distortions in the ring geometry, no apparent ruptures in the wheel structures are seen, pointing to effective tunnelling arising from conjugation in the rim.

It is not immediately obvious whether conjugation extends along the entire rim and whether the ring should be emissive at all. In molecules of six-fold symmetry, the $S_0$-$S_1$ transition is suppressed, as for example in benzene. Some larger macrocycles have also shown inhibited fundamental transitions[29-33]. However, slight interactions with the environment can break molecular symmetry, making the transition allowed. A notable example is the B850 band in the light-harvesting system LHII: based on dipole selection rules, thermally-activated emission from B850 would be expected since the lowest-lying state is dipole-forbidden[34]. However, thermally-activated emission is not observed experimentally[34].

**1** is indeed highly emissive, with a quantum yield of 71±5% and a short radiative lifetime of 840±60ps. The results of room-temperature absorption and emission spectroscopy (see *Supplementary Information*) of the ring and model linear compounds in solution are summarised in Table 1. The oligomers exhibit a bathochromic shift in emission and absorption with increasing size from the monomer to the dimer, implying improved electronic delocalisation[35]. Little difference is seen between dimer and hexamer, illustrating that delocalisation does not extend significantly beyond two monomers. Comparable porphyrin-based conjugated ring structures[21] fluoresce in the near-IR with a quantum yield of 0.12%, nearly three orders of magnitude lower than found here.

*Room-temperature single-molecule spectroscopy*

What is the microscopic nature of absorption and emission in **1**? Which part of the ring absorbs light, where is light emitted? Figure 2a sketches the problem. In excitation, ring symmetry should be preserved: incident light of *any* polarisation can be absorbed. Yet emission should arise from the formation of a linear transition dipole, situated anywhere on the annulus. Slight structural relaxation in the excited state or perturbation of the structure by the environment will break symmetry[34,36-38], generating distinct local potential minima[11] into which the exciton relaxes randomly. These microscopic characteristics of the π-system are best resolved by single-molecule techniques which overcome random averaging effects between molecules. Single molecules were diluted in Zeonex or poly(methyl-methacrylate) matrices at picomolar concentrations and imaged at room-temperature and 4K in two separate fluorescence microscopes. We first address the nature of the absorbing transition dipole by determining the distribution of excitation polarisation anisotropy values from molecule to molecule. Polarisation anisotropy is quantified in terms of linear dichroism. We recorded PL intensity under alternating horizontally (*H*) and vertically (*V*) polarised excitation. Linear dichroism is defined by the ratio of emission intensities *I* under the two excitation polarisations, $D_{excitation}=(I_V-I_H)/(I_V+I_H)$[4,39]. A linear dipole oriented randomly in a plane will yield a distribution peaking at $\pm 1$[40]. A value of $D_{excitation}=0$ will originate either from unpolarised absorption, or from a dipole oriented at 45° with respect to both excitation planes. Since molecule and dipole orientation are randomly distributed, the statistics of $D_{excitation}$ provide information on whether molecular absorption is polarised or unpolarised. Fig. 2b shows histograms of linear dichroism in excitation for the dimer, hexamer and ring. Following Table 1, the dimer constitutes the effective exciton size in hexamer and ring. As expected, the dimer displays an almost linearly-polarised absorption, the distribution matching a simple statistical simulation for a linear dipole (see Ref. 40). Deviations from a perfect linear dipole arise since the dimer is slightly bent, as seen in the STM image. As length increases, the conjugated system becomes more distorted, lowering the measured $D_{excitation}$ values and narrowing the distribution. For **1**, $D_{excitation}$ approaches zero: the molecules are effectively unpolarised absorbers. To assess the experimental resolution, a histogram of 200 fluorescent beads of comparable spectral properties and photon count rates – perfect unpolarised multichromophoric absorbers and emitters – is superimposed (black bars). Identical results are found in wide-field microscopy, implying that the confocal laser excitation is free of polarisation distortion artefacts despite the high numerical aperture objective used[41] (see *Supplementary Information* for discussion). Comparison of the rings to LHII is also interesting: despite similar dimensions and symmetry, isotropic absorption is *not* seen in experiments involving single complexes[36-38]. Other synthetic ring structures have only been studied by ensemble fluorescence depolarisation[15,21], revealing ultrafast loss in polarisation memory, compatible with our single-molecule results.

At the excitation wavelength used (405nm), the spokes of the wheel account for ~20% of overall absorption, since spoke and rim absorption overlap spectrally (see *Supplementary Figure S6*). It is only possible to discriminate spokes and rim in *emission*, where the spectra are shifted by 43nm. Exciton generation must lead to symmetry breaking in the excited state due to changes in bond-length alternation and structural relaxation[11]. We can image this symmetry breaking directly through the polarisation anisotropy in *emission*, which is again measured by the linear dichroism. Fluorescence is excited with alternating $H$ and $V$ polarised light, and the emission is split into two orthogonally-polarised components from which $D_{emission}$ is computed as illustrated on the right side of Fig. 2a (see *Supplementary Information* for details). $D_{excitation}$ and $D_{emission}$ can therefore be calculated for the same molecule, which makes the $D_{excitation}$ and $D_{emission}$ histograms directly comparable. Identical results are found for excitation with circularly polarised light. Only the first 100ms of illumination are considered for reasons discussed below. For the dimer, the $D_{emission}$ distribution is virtually identical to that in excitation since the molecule is too small for additional localisation to arise in the excited state. For the hexamer, the effect of exciton localisation to a unit comparable to the dimer is clearly visible: the $D_{emission}$ distribution resembles that of the dimer. More dipole orientations are available for excitation than for emission of the hexamer, making the $D_{excitation}$ histogram narrower than $D_{emission}$. The situation is very different for **1**. Although the photophysics is comparable to that of the dimer, the $D_{emission}$ histogram is much narrower than that of the dimer, implying that single molecules appear to primarily emit *unpolarised* light. Unpolarised reference beads show identical histograms of $D_{excitation}$ and $D_{emission}$ (black bars). However, for **1** the $D_{emission}$ distribution is not as narrow as $D_{excitation}$. This broadening of the $D_{emission}$ histogram implies the possibility of excited-state localisation occurring on some single rings during the experiment. In contrast, if localisation took place on *all* molecules within 100ms of illumination and always occurred *deterministically*, i.e. at the same position, the $D_{emission}$ histogram would match that of the dimer.

The dynamic nature of exciton-phonon coupling can give rise to localisation fluctuations and thus unpolarised PL, but so would simultaneous emission from multiple chromophores. To exclude this possibility, we study the statistics of single photons emitted by the rings. Figure 2c shows the photon correlation, measured by splitting the emission into two equal paths and recording photon arrival times as illustrated in the sketch, averaged over 100 single molecules. The dashed horizontal lines indicate the anticipated correlation thresholds based on photon count rates and background signal for one and two photons, respectively. Only one photon is emitted at a time, leading to the pronounced photon antibunching dip at zero delay between the two detectors. However, the polarisation of this photon is not predetermined. Panel d shows the temporal cross-correlation of two orthogonally-polarised detectors averaged over 95 molecules (first 100ms of illumination). The correlation is flat: there is no characteristic timescale for polarisation fluctuations. These results imply that single photons are emitted one at a time, and originate from different randomly-varying segments on the ring, leading to arbitrary fluctuations in emission polarisation. Molecular symmetry is broken spontaneously, and different localisation occurs following each photoexcitation event.

The situation becomes very different on longer timescales. Under prolonged illumination, photomodification of the molecule may occur, for example through the generation of a radical species[10]. This photomodification will lead to a *quasi-static* breaking of molecular symmetry, sketched in Fig. 3a. Changes in photomodification with time can induce fluctuations in excited-state localisation. This effect is clearly resolved in the temporal evolution of linear dichroism. Panel b exemplifies a molecule for which linear dichroism is measured *simultaneously* in excitation and emission (further examples are given in *Supplementary* Figure S7). The total fluorescence intensity shows discrete single-step blinking and an approximate halving of count rate after 12s of illumination. The corresponding $D_{excitation}$ is zero for the first 10s, and subsequently rises as the overall emission intensity decreases, presumably since part of the ring is photobleached[10]. $D_{emission}$ sets out at zero, drifting to a value of 0.4, and subsequently shows strong discrete jumps between positive and negative values. After continued illumination, fluctuations in localisation occur, leading to jumps in transition dipole orientation. The temporal evolution of linear dichroism due to photomodification can be visualised by plotting the $D_{emission}$ histogram as a function of time in panel c. We select molecules which exhibit $|D_{emission}| < 0.1$ within the first 100ms of illumination and do not drop by more than 30% in intensity over 5s. This initial $D_{emission}$-value can correspond to unpolarised PL or emission from a dipole at ~45° orientation with respect to the two detectors. Out of 2,000 single molecules, 32% fell within this narrow range. For a random distribution of dipoles, one would expect[40] only 6% of all molecules to be oriented at ~45°, implying that the low $D_{emission}$-values arise primarily due to unpolarised emission. The histogram clearly broadens over time. No such broadening is observed for dimers (not shown). We did not find evidence that this broadening is reversible under interruption of the illumination (see *Supplementary* Figure S8). No broadening is seen in the corresponding $D_{excitation}$ histogram, implying that the ring remains, to a first approximation, an isotropic absorber under continued illumination.

At the onset of photoexcitation, i.e. within the first ~100ms, single-ring fluorescence can appear *unpolarised*: the emission jumps between equally-weighted chromophores on the ring following every absorption event. With time, random photoinduced localisation occurs, leading to non-zero $D_{emission}$-values, which can then switch or drift randomly over milliseconds to seconds. Although such millisecond fluctuations in linear dichroism have previously been observed in multichromophoric macromolecules[40,42-43], their structural or electronic origin remains unclear. The rings demonstrate that these fluctuations are a *secondary effect* and only arise as a consequence of illumination. Initially, the symmetry of the molecule is preserved, so that random *spontaneous* symmetry breaking in the excited state can be observed. Consequently, spontaneous symmetry breaking is also virtually independent of temperature since it constitutes a purely intramolecular effect with no coupling to the heat bath: at 4K, ~20% of rings showed ($|D_{emission}| < 0.1$) under initial illumination, which is comparable to the room-temperature measurement (see *Supplementary Information*).

*Cryogenic single-molecule spectroscopy*

While not affecting spontaneous symmetry breaking, cryogenic temperatures do offer the advantage of overcoming thermal broadening to reveal the energetic heterogeneity of different chromophores within one ring molecule. In conjugated polymers, single chromophore spectra with linewidths orders of magnitude narrower than the ensemble have been identified[44-45]. Analogously, we can resolve typical[45] single-chromophore transitions in single rings at 4K, as described in Fig. 4a (black line). The spectrum consists of a strong, asymmetric peak at 465nm, followed by a series of similar but weaker vibronics. Whereas the single molecule exhibits a linewidth of <2nm, the ensemble solution spectrum in panel a (green line) spans >20nm. The histogram of 0-0 transitions of 117 different single molecules (grey bars) closely matches the 0-0 electronic transition in the ensemble (green line), implying that the ensemble is made up of distinct single-molecule transitions which differ due to varying interactions with the environment. Chromophores can therefore be distinguished by their transition energy[46]. Zero-phonon lines recorded above 500nm likely correspond to emission from ring aggregates since occurrence depends on molecular concentration.

Emission switching *between* chromophores in a single conjugated polymer chain is generally accompanied not only by a change in dipole orientation but also by a modification of transition energy[46]. This situation can also be observed in **1**, where we resolved linear dichroism *spectrally* by splitting emission into orthogonal polarisation components. To study the fluctuations in exciton localisation and its impact on the chromophore energy, we focus on molecules showing polarised emission following prolonged illumination (see *Supplementary Information* for further discussion). Figure 4b gives an example of a fluorescence spectral trace, resolving the two polarisation components (coloured red and blue). At 4K, fluctuations in linear dichroism are slower than at room temperature. Weak spectral jitter, characteristic of single-chromophore transitions[45], is visible in the trace. 120s into the measurement, the emission polarisation jumps, as does the transition wavelength, following a blinking event in which the molecule turned dark. This situation corresponds to dipole rotation by approximately 45° since only one polarisation channel is active before the event but both are equally strong afterwards, as seen in the integrated emission intensity plotted beneath. The case is very different for the molecule in panel c. Discrete reversible switching in dipole orientation occurs *without* discernible spectral change, leading to an anticorrelation of $H$ and $V$ components. Spectrally, it appears that only *one* chromophore is emitting, even though the emissive part of the ring rotates by ~90°. This observation suggests that the π-system experiences a homogeneous environment on length scales exceeding the size of the dimer, the emissive unit in the ring: even though the exciton localises to different parts of the molecule, in some situations, the same effective dielectric environment is probed so that the transition energy remains unchanged. This phenomenon could potentially arise due to long-range electronic correlations existing in the bath[47]. Such effects are, however, quite rare: out of 89 single molecules at 4K, 7 showed jumps in polarisation without a spectral shift, and 4 with one.

In conclusion, we have demonstrated that exciton localisation is a fundamentally non-deterministic process, arising randomly on different monomer units. The phenomenon is important to microscopic modelling of energy transfer pathways in organic electronic devices, and may contribute to the origin of intramolecular interchromophoric electronic coherences reported in conjugated polymers[2]: chromophores, the polarisable species, are not necessarily static entities.

**Acknowledgements**

The authors are indebted to the Volkswagen Foundation for providing generous collaborative funding. AT acknowledges financial support by the Fonds der Chemischen Industrie. JML is a David & Lucile Packard Foundation fellow and is grateful for an ERC Starting Grant (MolMesON, #305020). Correspondence and requests for materials should be addressed to J.M.L. or S.H.

**Author Contributions**

V.A., A.I., D.K. and S.H. designed and synthesised the compounds. A.T., D.W., T.S., F.S., J.V. and J. M. L. conceived, designed and performed the spectroscopy experiments and analysed the data. S. S. J. and S. H. performed and interpreted the STM experiments. A. T., J.V., S. H. and J. M. L. wrote the manuscript.

**Methods**

The synthetic methods and characterisation of the materials are described in the *Supplementary Information*.

Single-molecule emission was studied either at room temperature, in air, or under cryogenic conditions, *in vacuo*. For room-temperature measurements, the analyte molecules were embedded in a poly(methyl methacrylate) (PMMA, Mn = 46 kDa from Sigma Aldrich Co.) host matrix. The following steps were conducted: (i) Borosilicate glass cover slips were cleaned in a 2% Hellmanex III (Hellma Analytics) solution, followed by rinsing with MilliQ water. (ii) The glass cover slips were transferred into a UV-ozone cleaner (Novascan, PSD Pro Series UV) to bleach the glass cover slips from residual contaminant fluorescent molecules. (iii) The analyte was diluted in toluene to single-molecule concentration ($\sim 10^{-12}$ M) and mixed with a 1 w/w % PMMA/toluene solution. (iv) The analyte/PMMA/toluene solution was dynamically spin-coated at 2000 rpm onto the glass cover slips, which leads to a film thickness of about 50 nm with an average analyte density of 40 individual molecules in a range of $50 \times 50$ μm$^2$. (v) Single-molecule emission was studied under ambient conditions. It was found that under the latter conditions, and under exposure to dry nitrogen at room temperature, the single-molecule emission was substantially reduced, which is most likely due to the build-up of triplet excitons which can be quenched by molecular oxygen.

Fluorescence transients including the linear dichroism in excitation and emission of single ring molecules were recorded in a confocal fluorescence microscope. An inverted microscope (Olympus, IX71) with a high numerical objective (Olympus, APON 60XOTIRF,

NA=1.49) was used. The excitation source was a fibre-coupled diode laser (PicoQuant, LDH-C-405) with a wavelength of 405 nm in quasi-continuous wave mode (pulsed excitation, 20 MHz repetition rate). The excitation light was passed through a clean-up filter (AHF Analysentechnik, HC Laser Clean-up MaxDiode 405/10) and a Glan-Thompson polariser to provide linearly polarised excitation light. The polarisation of the excitation light was switched by an electro-optical modulator (FastPulse Technology Inc., 3079-4PW) and an additional $\lambda/4$ waveplate between horizontal and vertical polarisation every 500 µs, as described elsewhere.[2] The laser beam was expanded and collimated via a lens system and coupled into the oil immersion objective through the back port of the microscope and a dichroic mirror (AHF Analysentechnik RDC 405 nt). A diffraction limited spot was generated with an excitation power of 50 nW to ensure that all measurements are performed in the linear excitation regime, far below single-molecule saturation intensity. The fluorescence signal was either split by a polarising beam splitter (Thorlabs, CM1-PBS251) into two orthogonal polarising detection channels or by a 50/50 beam splitter into two equivalent detection channels. Avalanche photodiodes from Picoquant (τ-SPAD-20) were used as detectors and the signal was recorded via a time-correlated single photon counting (TCSPC) module from Picoquant (HydraHarp 400).

For low-temperature measurements, the ring molecules were dispersed in an optically inert polymer matrix (Zeonex 480, Zeon Corporation) at concentrations of $10^{-6}$ g/L and spin-coated on quartz substrates in a glove box under nitrogen atmosphere yielding film thicknesses of 20 nm. The samples were mounted on the cold finger of a He cryostat (ST-500, Janis Research Company Inc.) and kept under a vacuum of $10^{-7}$ mbar during the measurement at 4 K. Fluorescence was detected with a long working-distance microscope objective (7.7 mm, NA 0.55, Olympus America Inc.) that projects the emission onto the entrance slit of a 50 cm spectrograph (ARC-1-015-500, Princeton Instruments) with a CCD camera (CoolSnap:HQ2, Princeton Instruments).

Full details of the experimental methods are disclosed in the Supplementary Information.

**Figure Captions**

**Figure 1. Structure and synthesis of the spokes and ring 1, and of the non-cyclic analogues 9-11.** (a) tBu-XPhos, Pd$_2$dba$_3$, NaOtBu, toluene, 80 °C, 1 h, 66 %; (b) K$_2$CO$_3$, THF, methanol, r.t., 1 h, 95 %; (c) Pd$_2$dba$_3$, PtBu$_3$, CuI, piperidine, 120 °C, 16 min (μW), 74 %; (d) TBAF, THF, r.t., 3 h, 74 %; (e) Pd(PPh$_3$)$_2$Cl$_2$, CuI, I$_2$, air, THF, NHiPr$_2$, 50 °C, 60 h, 54 %.

**Table 1. Absorption, photoluminescence (PL), photoluminescence quantum yield (PLQY) and PL lifetimes measured in toluene or chloroform solutions at room temperature.**

**Figure 2. Fluctuations in exciton localisation due to spontaneous symmetry breaking.** a) Excitation with arbitrarily-polarised light leads to exciton formation. Excitons relax to segments of length comparable to the dimer. Due to bond-length changes in the excited state, the local potential in the proximity of the exciton is modified. Such exciton localisation occurs anywhere on the ring. Linear dichroism can be measured in either excitation (by switching laser polarisation, left) or emission (by passing fluorescence through a polarising beam splitter, right). b) Linear dichroism histograms in excitation and emission for 1597 dimer, 1273 hexamer and 730 ring molecules. Black bars indicate instrument response for fluorescent beads. STM images are shown with the graphite substrate axes indicated in white. c) Temporal photon correlation (pulsed excitation). Fluorescence passes through a beam splitter and is recorded with two photodiodes. At delay $\tau=0$ns between the detectors, photon coincidence approaches zero: photon antibunching implies the activity of precisely one chromophore. Dashed lines show calculated thresholds for one and two emitters. The experiment was carried out at 20MHz repetition rate, giving 50ns spacing on the $\tau$-axis. d) Cross correlation $g^2_{\text{cross}}(\tau)$ between two detector channels of orthogonal polarisation showing no discernible time scale for polarisation fluctuations, implying that ring emission initially appears unpolarised.

**Figure 3. Photoinduced fluctuations in exciton localisation apparent in the temporal dynamics of single-ring luminescence.** a) Photomodification of the ring can distort the excited-state potential quasi-statically, leading to a preferred polarisation in emission. Changes in photomodification will result in switching of emission polarisation. b) PL intensity, excitation and emission anisotropy of a single ring molecule. At short times into the measurement, the emission polarisation appears isotropic. The excitation anisotropy remains zero while the emission anisotropy initially increases from zero and subsequently exhibits random jumps, corresponding to changes in emissive dipole orientation. c) Evolution of the linear dichroism histogram in excitation and emission with time for 644 single ring molecules. The excitation remains isotropic whereas the emission becomes anisotropic with time due to photoinduced localisation of the excited state.

**Figure 4. Low-temperature PL spectroscopy of single rings showing switching in transition energy and polarisation.** a) A typical single-molecule PL spectrum at 4K

(black line), exhibiting a dominant zero-phonon line and discrete vibronic side bands, and the inhomogeneously-broadened ensemble solution spectrum at 300K (green line). The distribution of zero-phonon transition wavelengths for 117 single rings is superimposed. b), c) Fluorescence spectral trace and intensity as a function of time resolved for horizontally (red) and vertically polarised (blue) PL. Jumps in polarisation generally coincide with a change in emission wavelength since a different chromophore emits on the ring (left at time 120s). Reversible switching of the polarisation can also occur without any change in emission energy (right), implying that the energy of different chromophores on the ring remains controlled by the same dielectric environment. Arrows mark the anticorrelation in reversible switching between polarisation planes.

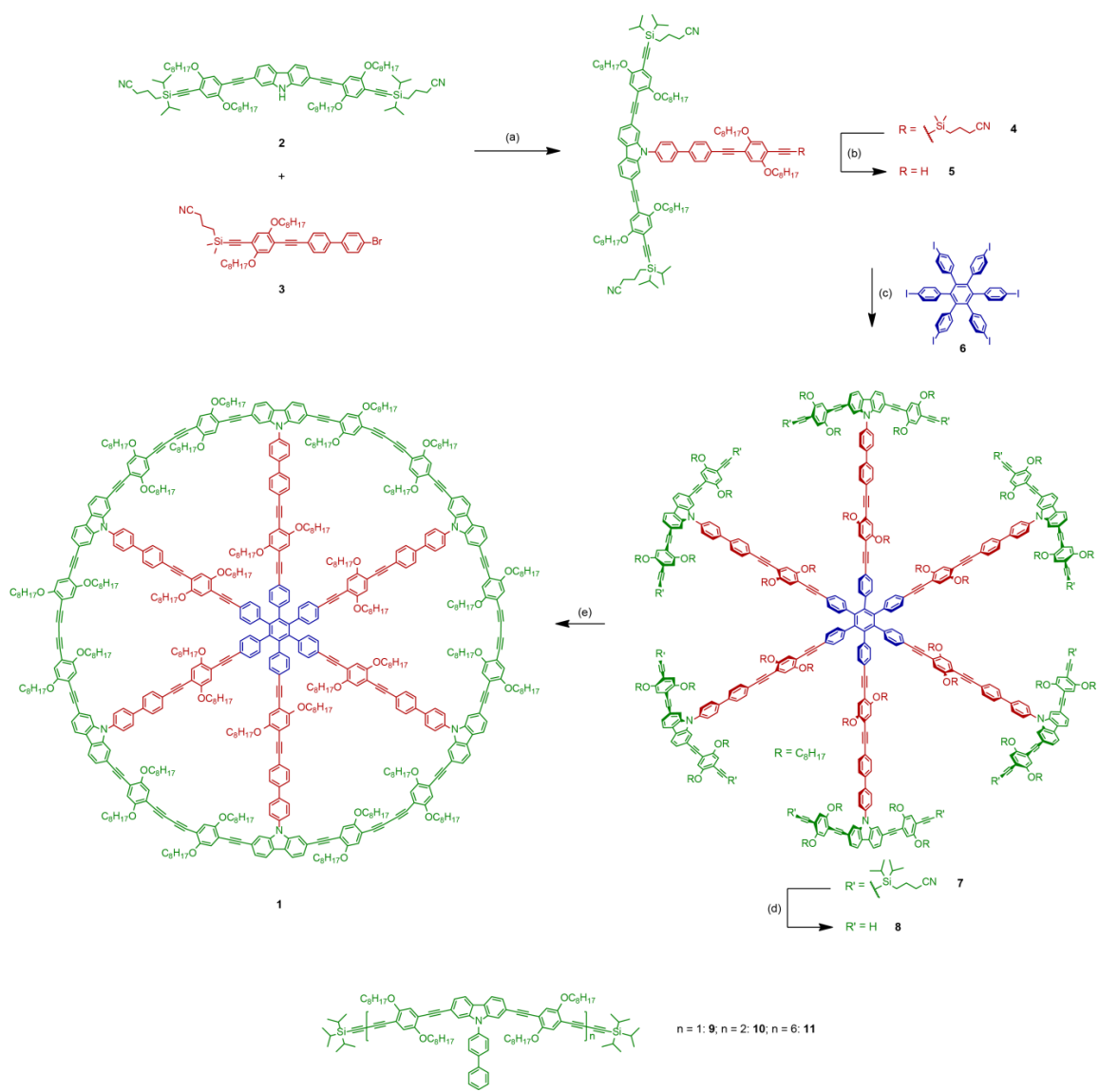

Figure 1.

|  | Absorption maximum (nm) | Molar absorptivity[a] ($cm^{-1}$ $M^{-1}$) [@$\lambda_{max}$ (nm)] | Absorption edge (nm) | PL maximum (nm) | PL lifetime (ns) | PLQY (%) | Radiative lifetime (ns) |
| --- | --- | --- | --- | --- | --- | --- | --- |
| Monomer **9** | 413 | 61,200 [417] | 431 | 431 | 0.97 ± 0.01 | 84 ± 5 | 1.15 ± 0.07 |
| Dimer **10** | 422 | 200,000 [425] | 459 | 459 | 0.58 ± 0.01 | 69 ± 5 | 0.84 ± 0.06 |
| Hexamer **11** | 426 | 500,000 [430] | 460 | 460 | 0.47 ± 0.01 | 66 ± 5 | 0.71 ± 0.06 |
| Ring **1** | 443 | 420,000 [444] | 462 | 462 | 0.60 ± 0.01 | 71 ± 5 | 0.84 ± 0.06 |

[a] at the maximum of the lowest energy absorption band in chloroform solution.

Table 1.

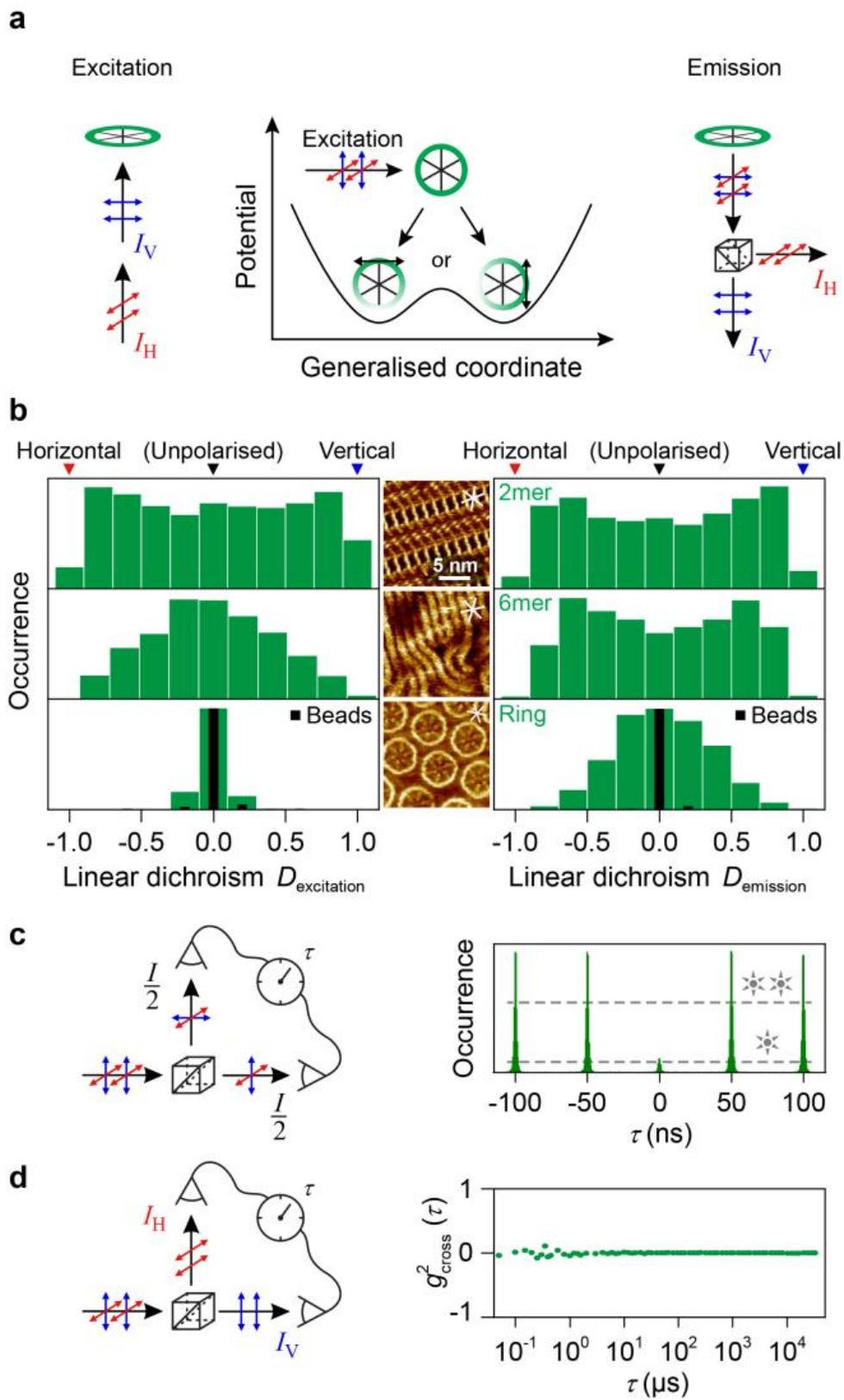

Figure 1.

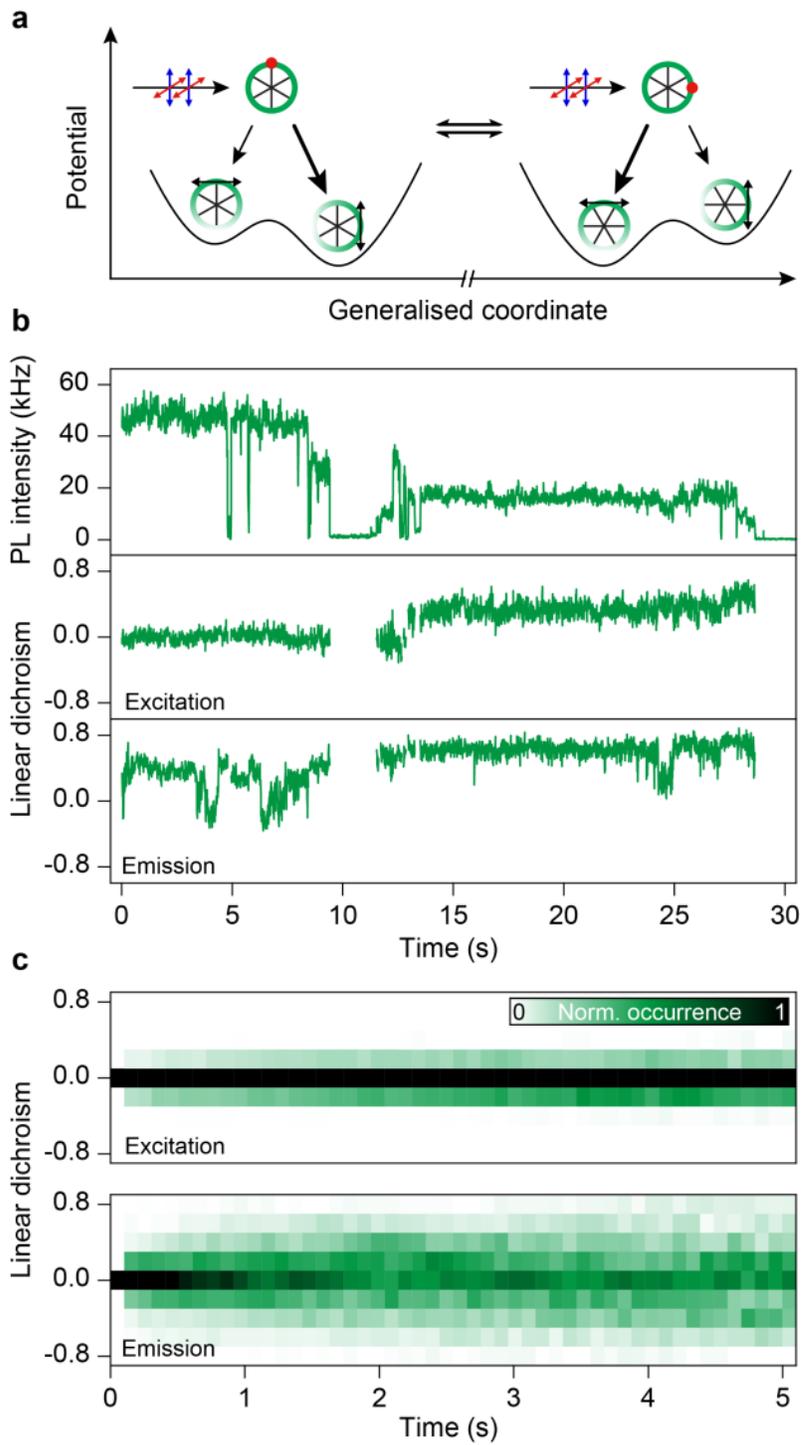

Figure 2.

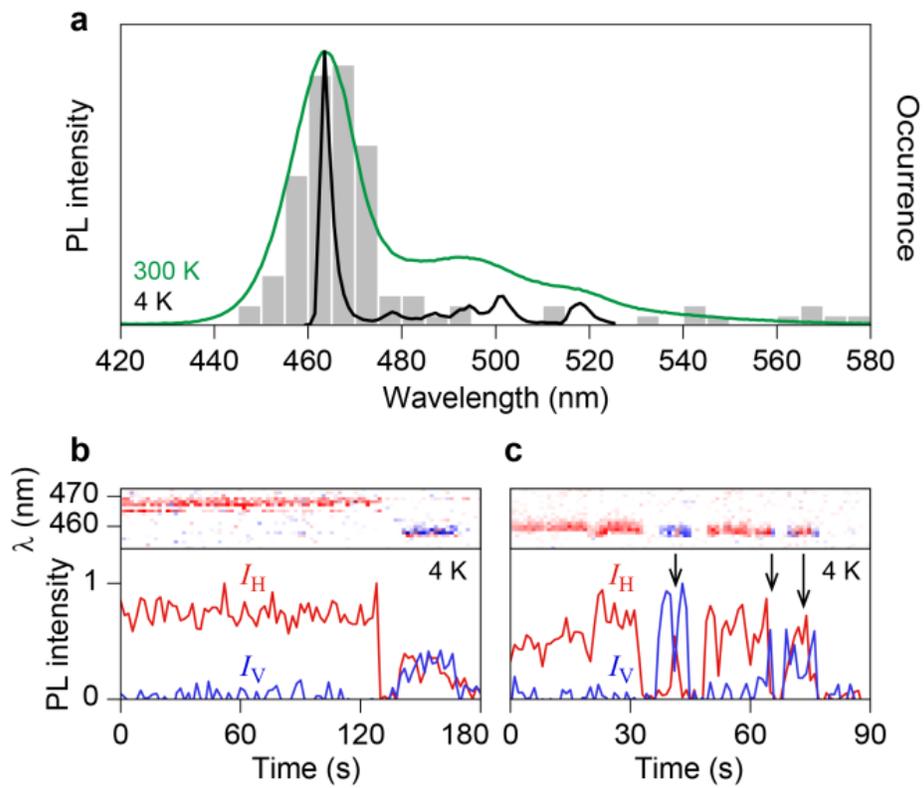

Figure 3.